\documentstyle[12pt]{l-aa}
\input{psfig}

\def\Real{{\rm I\mathchoice{\kern-0.70mm}{\kern-0.70mm}{\kern-0.65mm}%
  {\kern-0.50mm}R}}

\font \bolditalics = cmmib10
\def\bx#1{\leavevmode\thinspace\hbox{\vrule\vtop{\vbox{\hrule\kern1pt
        \hbox{\vphantom{\tt/}\thinspace{\bf#1}\thinspace}}
      \kern1pt\hrule}\vrule}\thinspace}

\def \vc #1{{\textfont1=\bolditalics \hbox{$\bf#1$}}}

\def\thetag{{\vc \theta}}

\def\be{\begin{equation}}
\def\ee{\end{equation}}
\def\ba{\begin{eqnarray}}
\def\ea{\end{eqnarray}}

\begin{document}


   \thesaurus{02         
              (12.03.4;
               12.04.1;
               12.07.1;
               12.12.1)}


   \title{Cosmic Shear Analysis in 50 Uncorrelated VLT Fields.
Implications for $\Omega_0$, $\sigma_8$
\thanks{Based on observations obtained at the
 Very Large Telescope UT1 (ANTU) which is operated by the
European Southern Observatory (program 63.O-0039A). }}
   \author{R. Maoli$^{1,2,3}$L. Van Waerbeke$^{1,4}$, Y. Mellier$^{1,2}$, 
 P. Schneider$^5$, B. Jain$^{6}$, F. Bernardeau$^7$, T. Erben$^{8,1,2}$, 
B. Fort$^1$ }
   \offprints{mellier@iap.fr}

  \institute{
   $^1$ Institut d'Astrophysique de Paris. 98 bis, boulevard
Arago. 75014 Paris, France. \\
$^2$ Observatoire de Paris. DEMIRM. 61, avenue de 
l'Observatoire.  75014 Paris, France.\\
$^3$ Universit\`a ``La Sapienza'' P.le Aldo Moro 2, 00185 Roma, Italy.
\\
$^4$ Canadian Institute for Theoretical Astrophysics, 60 St 
George Street, Toronto, M5S 3H8 Ontario, Canada.\\
$^5$ Institut f\"ur Astrophysik und Extraterrestrische Forschung 
der Universit\"at Bonn , Auf dem H\"ugel 71, D-53121 Bonn, Germany.\\
$^6$ Dept. of Physics, Johns Hopkins University, Baltimore, MD 21218,
USA\\
   $^7$ Service de Physique Th\'eorique. C.E. de Saclay. 91191 Gif sur 
Yvette
Cedex, France.\\
   $^8$ Max Planck Institut fur Astrophysiks, Karl-Schwarzschild-Str. 1,
Postfach 1523,
D-85740 Garching, Germany. \\
}



\maketitle
   \markboth{Cosmic shear with the VLT}{Maoli et al.}

\begin{abstract}
We observed with the camera FORS1 on the 
VLT (UT1, ANTU) 50 randomly selected fields
and analyzed the cosmic shear inside 
circular apertures with diameter ranging from 0.5 to 5.0 
arc-minutes. The images were obtained in
optimal conditions by using the Service Observing 
proposed by ESO on the VLT, which enabled us to get 
a well-defined and homogeneous set of  data.  The 50 fields 
cover a 0.65 square-degrees area
spread over more than 1000 square-degrees which provides a sample 
ideal for minimizing the cosmic variance.  Using the same techniques as
in \cite{VW00}, we measured the cosmic shear
signal and investigated the systematics of the VLT sample. 
We find a significant excess of correlations between galaxy 
ellipticities on those angular scales.  The amplitude and the 
shape of the correlation as function of angular scale are remarkably
similar to those reported so far.  \\
Using our combined VLT and CFHT data and adding the results published 
by other teams we put the first joint constraints
on $\Omega_0$ and $\sigma_8$ using cosmic shear surveys.
From a deduced average of the redshift of the sources
the combined data are consistent with
$\sigma_8\simeq 0.59 ^{+0.03}_{-0.03} \Omega_0^{-0.47}$
(for a CDM power spectrum and $\Gamma=0.21$),
in excellent agreement with the cosmological 
constraints obtained from the local cluster abundance.

\keywords{Cosmology: theory, dark matter, gravitational lenses, large-scale
structure of the universe}
\end{abstract}


\section{Introduction}
The weak lensing by large scale structures (hereafter cosmic shear)
probes the projected mass density 
along the line-of-sight, regardless of the distribution of light. 
This could yield powerful constraints on
the evolution  of the dark matter power spectrum,
the cosmological parameters and the mass/light bias
as a function of the angular scale and the redshift
(see reviews from \cite{M99}, \cite{BS99} and references therein).

Four teams announced recently the 
first measurements of a significant variance of the shear $\langle
\gamma^2\rangle$ on scales below 
30 arc-minutes (\cite{VW00}, \cite{BRE00}, \cite{Witt00}, \cite{K00}).
All the measurements are in remarkable agreement, and it is worth to
mention that these analyses were 
done from independent data sets obtained on various telescopes
and by measuring galaxy shapes with different tools.
Moreover the signal is consistent with the theoretical expectations of    
current cosmological models, and does not show 
 strong contamination by residual
systematics. Though it is not yet possible to break the 
degeneracy between the normalization of the power spectrum, $\sigma_8$,
and the density parameter $\Omega_0$ from a measure of
$\langle\gamma^2\rangle$ alone (\cite{B97},\cite{jain97}), some
cosmological models can already be rejected
(like the COBE-normalized SCDM which predicts too much power at small scale,
at the 5-$\sigma$ level).
These encouraging results show that cosmic shear
is close to be a mature tool and offers a neat complementary
approach in cosmology compared to CMB anisotropies, SNIa,
or galaxy catalogue statistics.

However, more quantitative and reliable constraints on 
cosmological models are still under way.  
The need for a high confidence level  
on the  cosmic shear signal demands an
excellent control of the systematics  and 
 a significant reduction of the cosmic variance (\cite{VW99}).   
 The former issue is discussed at length by \cite{VW00} and 
 \cite{BRE00} on real data, and it is also addressed in
\cite{Ks00}, \cite{E00} and \cite{BRCE00} using simulated data.
 They have carried out convincing tests and simulations in order to 
 demonstrate that gravitational weak distortion is measurable 
 down to slightly below the percent level, without critical issues regarding 
  residuals from the PSF corrections.

In order to convince that the signal is not
contaminated by some unexpected systematics, it is still important
to diversify the observing conditions, and produce
more independent data sets with different systematics.
Of equal importance,  the field-to-field variations, caused by source
and lens clusterings and by cosmic variance,
are an additional source of noise.
The four studies mentionned above cover a total area  
of about 5 deg$^2$ but only  totalize 28 {\it uncorrelated} fields.   
Clearly, one needs to increase first the number of uncorrelated
fields, rather than the total solid angle of the survey, 
to explore the field-to-field variation of the 
cosmic shear amplitude and to minimize the 
cosmic variance.  

This work explores a new data set obtained with the faint 
object spectrograph FORS1 (\cite{Appen98})  mounted on the first VLT (UT1/ANTU) 
at Paranal (program 63.O-0039A; PI: Mellier).  These data provide  
independent measurements 
of the cosmic shear signal, in addition  to the four previous studies,
but with a totally new technology telescope and a new observing
mode. In contrast to previous telescopes
the VLT  is continuously operated in an active optics mode and 
offers the possibility to observe in service mode.
Thanks to its large collecting power,  
the VLT can observe many single fields very quickly
which permits to do deep  pointings on many different lines-of-sights. This 
new strategy, complementary to the one we adopted on the CFHT data 
(\cite{VW00}), gave us an interesting large complementary sample 
of 50  VLT/FORS1 uncorrelated fields for cosmic shear.  The analysis of
these fields is discussed below. They are then combined with other existing
results to give the first constraints on ($\Omega_0$,$\sigma_8$) derived
from cosmic shear.

The paper is organized as follows: Section 2 is a description of
the data and the observations. Details on data reduction are given
in Section 3. The analysis of the fields and 
the measurements of the shapes of galaxies are developed in
Section 4.  In Section 5 we discuss the systematics. 
Discussion and combination with other results in a cosmological context is
given in Section 6.

\section{Data and Observations}
The targets were selected from three very large sky areas 
close to the North and South galactic poles with the following criteria:
\begin{enumerate}
\item no stars brighter than 8th magnitude inside a circle of 1 degree around
the FORS field (to avoid light scattering);
\item no stars brighter than 14th magnitude inside the FORS field;
\item no extended  bright galaxies in the field. Their luminous 
halo could contaminate
the shape of galaxy located around them; 
\item no rejection of over-dense regions, where clusters or groups of 
galaxies could be present
from visual inspection of the Digital Sky Survey, and no preferences
towards empty fields. This criterion avoids biases toward under-dense
regions with systematically low value of the convergence;
\item minimum separation between each pointing of at least 5 degrees in
order to minimize the correlation between fields;
\item galactic latitudes lower than $70^{\circ}$ in order to 
get  enough stars per field  for the PSF correction.
\end{enumerate}
We selected 50 FORS1 fields, each covering 6.8'$\times$6.8'. 
The total field of view is 0.64 deg$^2$ and the pointings are randomly 
distributed over  more than 1000 deg$^2$. Among those 50 fields, 
two have been selected from the STIS parallel data sample, for 
cosmic shear analysis on very small scale, and two are in common with the WHT 
sample observed by \cite{BRE00}
\footnote{We plan to check the reliability of 
the measurements by comparing the analyses done by each group on 
these common fields}.
  
The observations are in  I-band data only and were obtained 
with FORS1 on the VLT/UT1 (ANTU) at the Paranal Observatory.  
They were carried out in service mode by the ESO staff during Period 63 
(March 1999 to September 1999) with the standard imaging mode of the 
spectrograph.  It turns out that 
our program is perfectly suited for service mode, both
from a  practical point of view, since they are easily set up  
during service mode periods on UT1, and from a  scientific 
point of view, since we have a guarantee to get a complete set
of  data within our specifications. \\
FORS1 is equipped with 
a 2048$\times$2048
thinned CCD, backside illuminated.  The pixel size is 24 $\mu$m, 
providing a scale of 0.2 arc-second in  standard mode of the 
instrument\footnote{see http://www.eso.org/instruments/fors1/index.html}.  
Remarkably, 
no fringe pattern has been detected on FORS1 with this filter,  so
I-band observations enable one to get optimal image quality on
each field.
Since we only requested  non-photometric nights, with seeing lower than 
0.8 arc-seconds, the schedule of our 50 fields over the semester 
was indeed quite easy.  
The total exposure time (36 minutes) was computed in order to reach 
$I=24.5$, which  corresponds to a galaxy number density per fields of about 
30 gal ${\rm arcmin}^{-2}$. The expected average redshift of the
lensed sources is $\langle z\rangle \approx 1$. Each pointing was split into 
at least $6\times 6$ minutes exposures with a random offset
of 10 arc-seconds in  between.
Some details  about the  50 fields are given in Table \ref{SumTab}. 

\section{Data reduction}
The data were received by the end of November 1999 at IAP.  The  processing 
of the 50 fields was done using the 
TERAPIX\footnote{see http://terapix.iap.fr} data center facilities. 
The FORS1 images are much smaller than the big CCD mosaics 
usually processed by TERAPIX, so we used a specific pipeline which includes 
some IRAF scripts. Since the FORS1 detector has four outputs, we 
simply consider each quadrant as an individual device 
during all the preprocessing stage. 

The master bias was built using a standard median filter.
For the superflat, since the CCD images do not 
show fringe patterns, the processing can be done in one step with
a simple un-smoothed master-flat  
to correct for the relative pixel-to-pixel variations.    
However, the multiplicity of observations done in service mode 
provides images taken during different Moon 
periods and with different weather conditions which cannot be assembled 
to build up a single superflat. In our case, when all the 
images are used, the global master-flat shows a strong and
persistent gradient along the CCD ($>10\%$).
It is therefore preferable to use  daily master flats by averaging 
only  the images taken during the same night.   The drawback 
is the small number of images which can be combined together.
If it is too small,  each individual frame 
has a strong relative weight which may results in  correlated noise between 
individual images and the master-flat. Therefore we decided to 
produce a special daily master-flat for each image:  each individual
frame was flattened by using a master-flat done with the rest 
of the data taken during the night.

The four quadrants are then put together and rescaled according to 
the gain of each output. If the sky backgrounds of the four quadrants
are still different, we proceed to another small rescaling 
accordingly. This procedure turns out to work well in most 
cases. However, some images are not perfectly flatfielded and still  show a 
small cross-shape discontinuity in the middle of the 
FORS1 images,  between the borders of the quadrants. 
This area has been masked in the final co-added images when necessary.

\begin{figure}
\centerline{
\psfig{figure=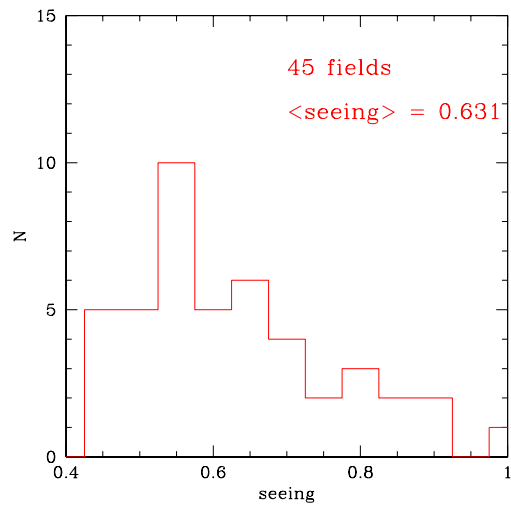,width=7cm}}
\caption{\label{seeing} Seeing distribution of the final VLT field sample 
  used to measure the cosmic shear.
}
\end{figure}

The stacking of each frame is done using standard IRAF procedure
. The final error for alignment 
 was less than one tenth of pixel.
The final set of 50 stacked fields turns out to have a remarkable seeing 
distribution (see Figure \ref{seeing}). In fact, 90\% of them are within our 
seeing specifications, with an average value of 0.63''.

\begin{figure}
\centerline{
\psfig{figure=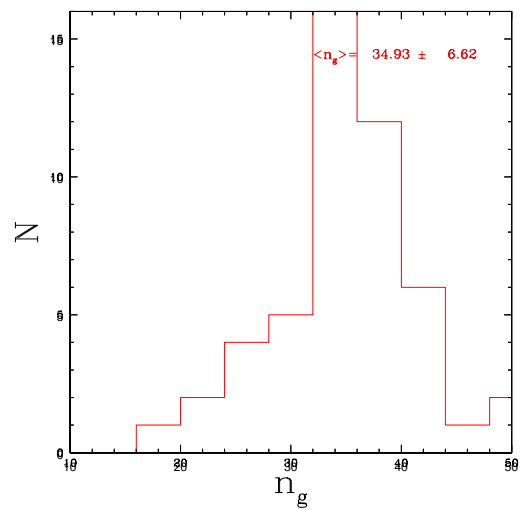,width=7cm}}
\caption{\label{ngal} Histogram of the galaxy number density distribution
in the VLT fields. The distribution shows a narrow peak at $n=35 \
{\rm gal~arcmin}^{-2}$.
}
\end{figure}
 
\begin{table*}
\caption{List of the 50 VLT/FORS1 fields. All have an
exposure time of at least 36 minutes and cover a total area of 0.64 deg$^2$.
  The seeing given in this table correspond to the stacked image.
}

\label{SumTab}
\begin{center}
\begin{tabular}{|l|c|c|c|c|l|}
\hline
Target Name &RA (J2000) & DEC (J2000) & Seeing& $N_{gal}$ & $I_{AB}$ Lim.\\
\hline
vlt27 &00 59 28.1 &-00 18 28 &0.72" & 991 & 24.5 \\
vlt28 &01 31 40.3 &-00 22 28 &0.54" & 1580 & 24.9   \\
vlt29 &01 59 40.8 &-00 03 51 &0.49" & 1878 & 25.1  \\
vlt30 &02 28 44.0 &-00 03 26 &0.54" & 1680 & 24.9  \\
vlt31 &01 00 21.8 &-03 15 31 &0.57" & 1539& 25.0 \\
vlt33 &02 00 08.1 &-03 00 30 &0.44" & 1755& 24.9 \\
vlt35 &00 59 35.3 &-06 10 05 &0.73" & 1201& 24.8 \\
vlt36 &01 28 53.1 &-06 01 39 &0.68" & 1595& 24.8 \\
vlt37 &01 57 05.8 &-06 05 01 &0.90" & 1138& 24.5 \\
vlt39 &21 30 45.3 &-09 58 45 &0.76" & 1369& 24.8 \\
vlt40 &22 04 37.9 &-10 15 09 &0.71" & 1454& 24.9 \\
vlt42 &22 29 29.2 &-10 12 01 &0.72" & 1268& 24.7 \\
vlt43 &21 30 25.3 &-15 11 48 &0.55" & 1525& 25.0 \\
vlt44 &22 02 16.6 &-14 53 03 &0.64" & 1715& 24.9\\
vlt45 &22 30 41.8 &-14 54 55 &0.46" & 1791 & 24.8\\
vlt46 &22 01 42.2 &-20 10 55 &0.65" & 1651& 25.0\\
vlt47 &22 29 33.8 &-20 14 44 &0.51" & 1639& 25.0\\
vlt48 &21 30 45.3 &-24 53 40 &0.63" & 1604& 24.8\\
vlt49 &21 58 44.7 &-24 57 15 &0.62" & 1494& 24.6\\
vlt50 &22 30 43.8 &-25 01 42 &0.55" & 1595& 25.0\\
vlt51 &20 59 30.5 &-30 18 31 &0.62" & 1243& 24.5\\
vlt52 &22 00 26.2 &-30 01 45 &0.65" & 1357& 24.8\\
vlt53 &22 31 15.3 &-30 07 15 &0.55" & 1584& 24.7 \\
vlt54 &21 29 53.6 &-34 51 52 &0.57" & 1629& 25.1\\
vlt55 &22 00 14.1 &-35 30 54 &0.53" & 1444 & 24.6\\
vlt56 &22 30 06.4 &-35 10 33 &0.83" & 1007& 24.5\\
vlt57 &21 28 04.9 &-39 49 02 &0.55" & 1650& 25.0\\
vlt58 &22 00 06.7 &-40 04 55 &0.49" & 1664& 24.8\\
vlt59 &22 29 11.8 &-39 36 28 &0.70" & 1453& 24.7\\
vlt60 &22 59 24.4 &-10 01 29 &0.47" & 1694& 25.0\\
vlt61 &22 59 24.2 &-15 08 47 &0.47" & 2115& 25.0\\
vlt62 &22 59 01.8 &-19 44 03 &0.47" & 1769& 25.0\\
vlt63 &22 59 39.5 &-24 52 51 &0.49" & 1502& 25.0\\
vlt64 &22 59 56.1 &-30 14 27 &0.60" & 1285& 24.5\\
vlt65 &23 00 44.3 &-34 55 26 &0.54" & 1675& 25.0\\
vlt66 &23 01 24.8 &-40 25 20 &0.77" & 896& 23.6 \\
vlt75 &21 28 14.7 &-20 07 18 &0.56" & 1683& 24.9\\
vlt76 &21 32 21.0 &-30 25 57 &0.63" & 1434& 24.6\\
vlt77 &14 59 07.4 & 00 07 54 &0.80" & 1414& 24.8 \\
vlt78 &14 59 03.2 & 05 11 32 &0.50" & 1862& 25.0\\
vlt79 &14 59 32.7 & 10 13 19 &0.76" & 1142& 25.0 \\
vlt80 &15 30 17.5 & 00 10 58 &0.60" & 1511& 24.8\\
vlt81 &15 29 40.4 & 04 54 10 &0.63" & 1405& 24.7\\
vlt82 &15 28 59.7 & 10 14 59 &0.59" & 1401& 24.6 \\
vlt83 &15 59 00.7 &-00 07 14 &0.87" & 1015& 24.3 \\
vlt84 &16 03 35.0 & 05 10 46 &0.91" & 923& 23.8\\
vlt85 &15 56 47.6 & 10 17 28 &0.66" & 1275& 24.7\\
vlt86 &16 00 30.1 & 14 58 35 &0.78" & 1231& 24.6\\
stis7new & 22 28 13.6 & -26 59 11 &0.76" & 860& 23.3\\
stis10new & 12 28 32.8& 02 10 05 &0.93" & 930& 24.0 \\
\hline
\end{tabular}
\end{center}
\end{table*}

The photometric calibrations of the FORS1 fields were done using the 
photometric standard star images provided by the ESO staff at
Paranal. The fields were  chosen in a fairly broad 
  sample of  the Landolt's catalog 
(\cite{landolt}):  SA92, SA95, SA101, SA109, SA110, MARK-A, 
  PG0131-051, PG1313-086, PG1633+099, PG2213-006, PG2331+055.  
The photometric zero-points 
were calculated using magnitudes computed by SExtractor
\footnote{Using MAG\_AUTO with a minimum radius of 3.5 pixels.} and by
rescaling the gains of each quadrant accordingly. We noticed a 
fluctuation of the zero-point of about $\pm 0.05$ magnitude along 
  the observing period, probably produced by the important variation of  
the atmospheric conditions during the whole semester.

The final photometric catalogs look sufficiently homogeneous from field to
field for our purpose. The magnitude histograms for each field
have the same shapes 
and show a cutoff at  $I_{AB}=24.5\ -\ 25.1$. This cut-off defines our
limiting magnitude.
Five fields have a brighter magnitude cut-off at $I_{AB}=23.5\ - 24.5$. These
fields have  the
worse observational condition: seeing higher than 0.8" and high sky background
due to the moonlight. Some fields have a higher number of co-added frames
but it does not change significantly the final magnitude limit of the 
50 fields.

\section{Measurement of the shear signal}
\subsection{Catalogue generation and selection criteria}

The galaxies have been processed using our modified version of 
IMCAT software\footnote{Made available by Nick 
Kaiser; http://www.ifa.hawaii.edu/$\sim$kaiser/}. The modifications
and the general analysis method have already been described in
Section 3 of \cite{VW00}.
Here we remind shortly the selection procedure, paying attention
only to the differences with the previous CFHT data analysis.

\begin{figure}
\centerline{
\psfig{figure=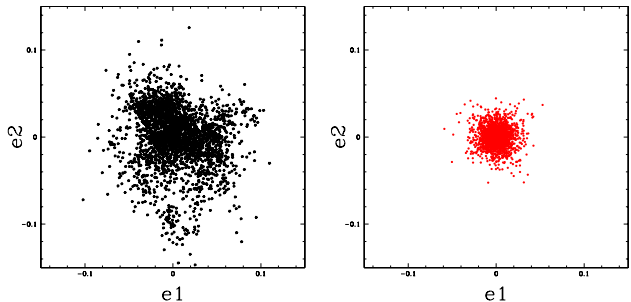,width=8.5cm}}
\caption{\label{refseeing} PSF ellipticity before (left) and after (right)
anisotropy correction.
}
\end{figure}

The principal steps are the followings:
\begin{itemize}
\item {\it Masking the boundaries:} the splitting procedure of  
exposures generates noisy thick strips 
surrounding the CCD,  with lower signal-to-noise ratio than the rest 
of the image, and a grid-like pattern produced by the 
stacking procedure.  We therefore mask automatically the  
boundaries of each image.
\item {\it Masking the central cross-shape area:} the four readout ports
of the CCD delineate four well-defined quadrants with different readout 
noise, charge transfer efficiency and gain. At the junction of the 
quadrants, the sharp changes of CCD properties appear as a cross-shape
area located in the middle of the CCD.  In principle, this 
area is very small, but, like at the boundaries, the small shifts between each
exposure enlarge its size  which thus 
appears as a broad region with smaller signal-to-noise ratio than the rest of
the image. On the fields affected by this effect, the central cross-area
is masked.
\item {\it Masking objects:} large areas around bright stars,   
saturated columns and diffraction patterns are 
masked on all images. They  can 
produce spurious elongated structures which can be misinterpreted as 
shear signal. Likewise, we also mask areas around  bright and/or 
extended galaxies, elongated 
stripes produced by residuals from asteroids, satellites or aircrafts.
The fraction of objects  removed by this masking procedure
(boundaries, cross-shape area and objects) varies
from 10\% to 20\%. 
\item {\it Stars selection:} we use a radius ($r_h$)
vs magnitude ($mag.$) plot to 
select unsaturated stars (we do not include stars closer than one
magnitude from the saturation level). The  vertical box which encompasses 
the stars is carefully sized  to avoid saturated stars and
the crowed faint-magnitude areas of 
the $r_h-mag.$ diagram where galaxies and stars mix together. 
This domain contains very low signal-to-noise objects and is most
of the time useless for the calibration.
\item {\it Fit of the stellar polarizability tensors:} using
these stars, we can map the PSF by fitting their raw shape 
to  a third order polynomial.  We then compute the quantities 
$\textstyle{p_\alpha={e_\alpha^\star\over P^{sm}_{\alpha\alpha}}}$ and 
$\textstyle{{P^{sh}_\star\over P^{sm}_\star}}$; 
\item {\it Source selection:} The starting sample of galaxies 
is extracted from the 50 VLT fields using SExtractor (\cite{BA96}) and
produces a raw catalogue of 72500 galaxies.  Objects at the edge of
the fields and deblended objects are
immediately removed. Galaxies with a half light radius smaller
than the largest size of the star box are also removed.
As pointed out by 
\cite{VW00}, close pairs of galaxies may produce spurious elongated 
galaxies, consequently pairs with separation $d < 10$ pixels are also removed.
We do not perform any magnitude selection, therefore the redshift distribution
is broad and can be described by:

\begin{equation}
p(z)=\beta {\Gamma((1+\alpha)/\beta)\over z_0} \left({z\over z_0}\right)^\alpha\exp{\left[-\left({z\over z_0}\right)^\beta\right]},
\label{z_source}
\end{equation}
with $\alpha=2, \beta=1.5, z_0=0.8$, which is consistent with a limiting
magnitude $I_{AB}\sim 24.5$ as given by \cite{cohenetal}.

\item {\it Estimate of the shear:} using the stellar fits, we compute the 
'pre-seeing' shear polarizability for all the selected sources as
described in \cite{VW00}.

Figure \ref{refseeing} shows the star ellipticities for all VLT fields 
before and after the PSF correction. 
One of the fields shows a seeing significantly alterated by strong wind
and its image quality is too far from our specifications, this field
(stis7new in Table \ref{SumTab}) was removed. We ended up with 49
fields with in average 30 stars/field corresponding to a total number 
of 46941 galaxies, that is $\sim 21 \ {\rm sources/arcmin^2}$. Note that this
is less than the number given in Table \ref{SumTab} which was the number
density {\it before} objects selection through our shape measurement process.

\end{itemize}

\subsection{Shear measurement}

At this stage, the variance of the shear can be computed exactly as 
described in \cite{VW00}. However here we have measured the signal in a
slightly different manner. Instead of computing the variance of the
shear by simply squaring the
averaged shear per cell, we directly removed the diagonal terms from the
squared quantity, such that the computed variance is by definition
an unbiased estimate of the true shear variance (see \cite{SvWJK98}).
Therefore we do not need to subtract the shot noise contribution as
in previous works, which was done using time-consuming monte-carlo
randomizations. Let us call $e_\alpha(\thetag_k)$ the
ellipticity of a galaxy at
a position $\thetag_k$, and $w_k$ its weight, calculated according to
\cite{VW00} (see Section 3.2 of that paper).
An unbiased estimate of the variance of the shear
$\gamma^2(\thetag_i)$ at location $\thetag_i$ can be directly obtained by
measuring:

%
%
\begin{equation}
E[\gamma^2(\thetag_i)]={\displaystyle \sum_{\alpha=1}^2 \sum_{k\ne l}^N w_k w_l
e_\alpha(\vec\theta_k) e_\alpha(\vec\theta_l)
\over
\displaystyle \sum_{k\ne l}^N w_k w_l}
\label{estimator}
\end{equation}

The final product, the variance of the corrected ellipticity of 
galaxies as function of angular smoothing scales, is summarized 
on Figure \ref{shearfigmodel}. On this Figure we also plot the results 
of the previous detections.   
The angular scale corresponds to the diameter of a circular 
top-hat filter (we rescaled accordingly the angular scales of the 
previous studies which were published for a squared top-hat).
The  1-$\sigma$ error bars have been calculated from 100 realizations of noisy
catalogues by randomizing the position angle of each galaxy.

The amplitude and the  shape of the signal are in good agreement with 
former results. They are within the 1-$\sigma$ error at 
all scales (note that the points are not independent). The 
agreement of these new measurements with previous ones provides a 
new independent indication that the signal is not produced by 
uncontrolled systematics. However, as for the CFHT analysis
(see \cite{VW00}), we must analyze carefully the systematics in the
VLT data.

\begin{figure}
\centerline{
\psfig{figure=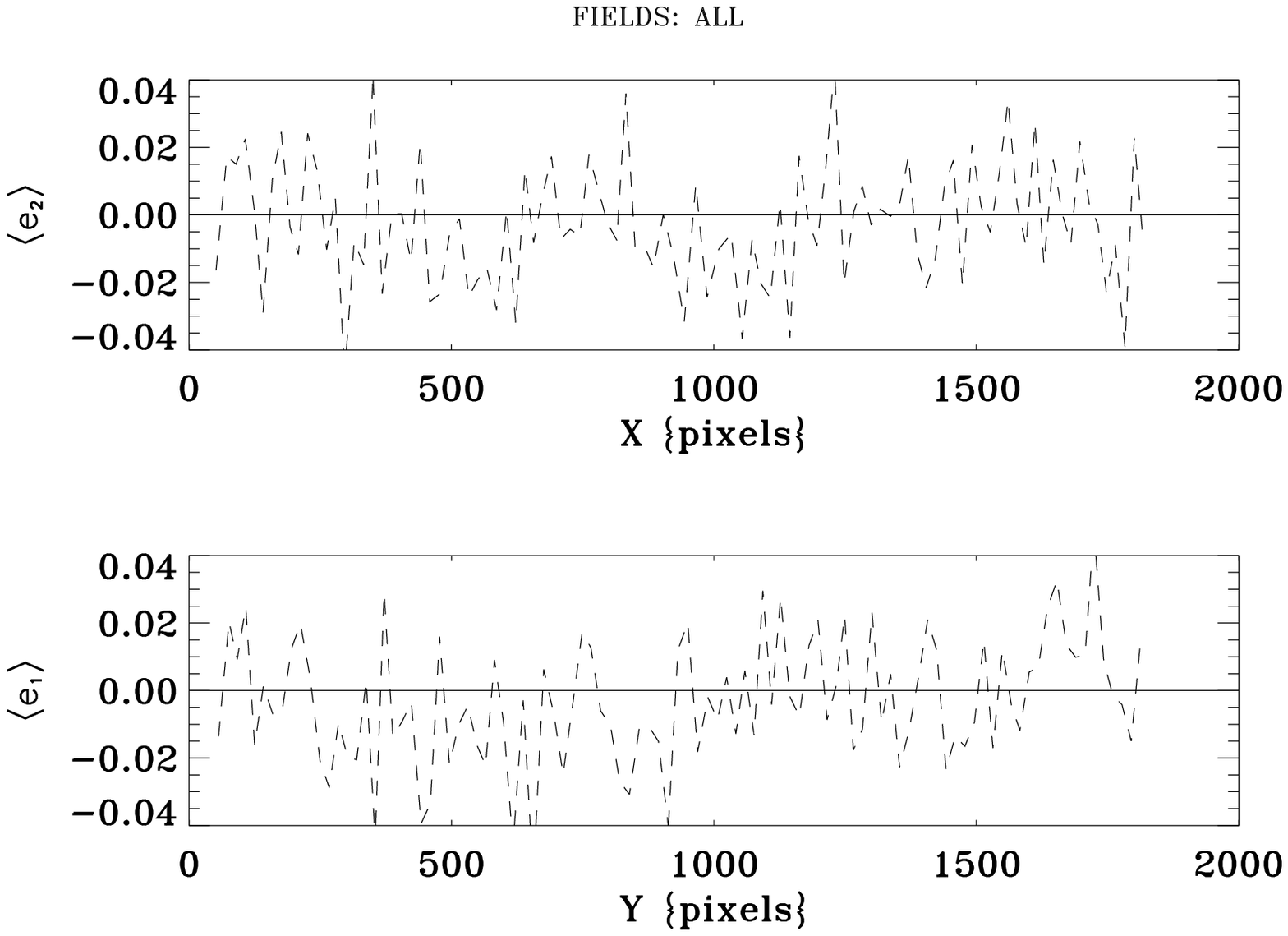,height=6cm}}
\centerline{
\psfig{figure=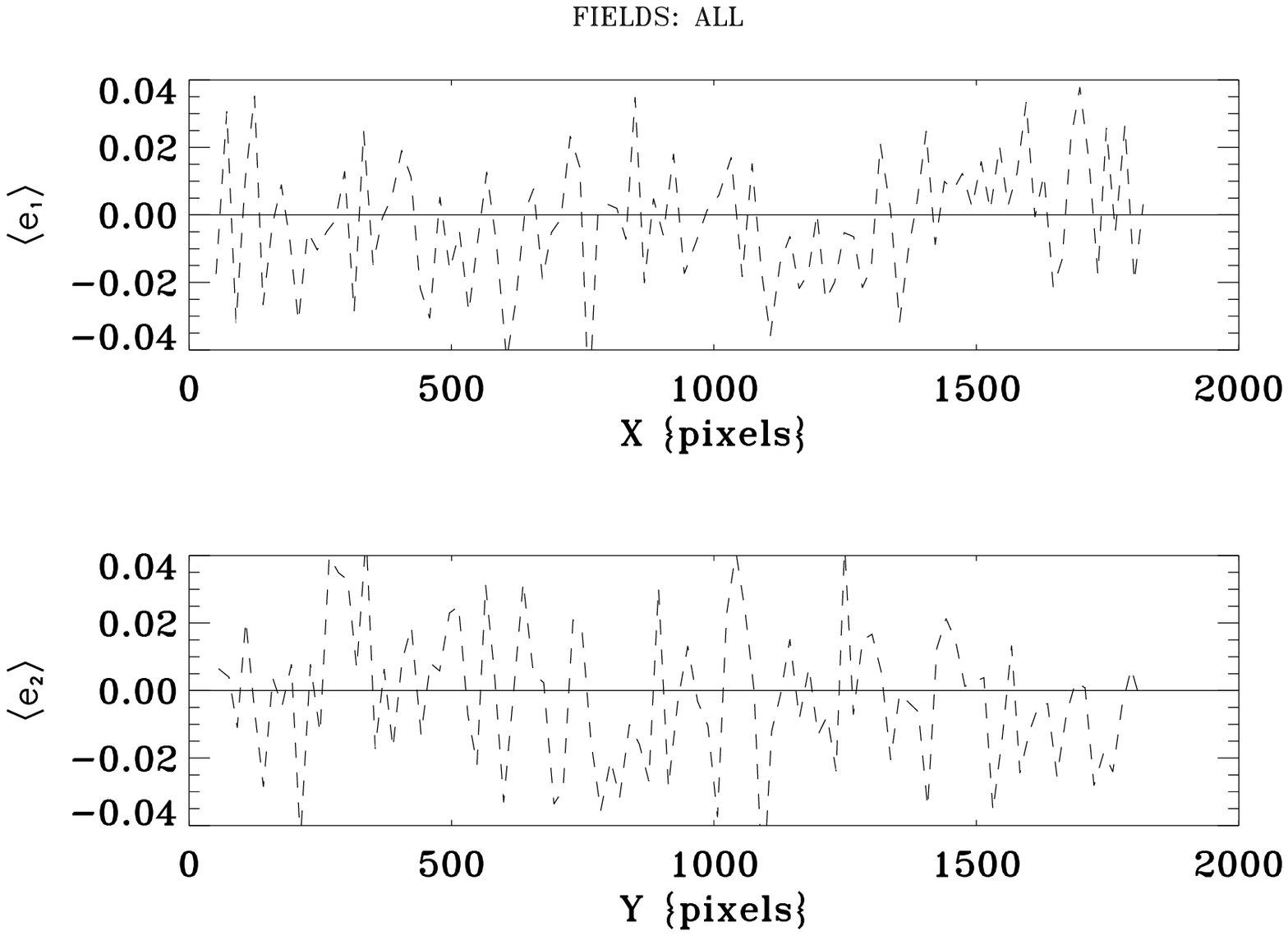,height=6cm}}
\caption{\label{crosseffect}  Average corrected galaxy ellipticity 
 $\left<e_{\alpha}\right>$ along vertical ($Y$) and horizontal
 ($X$)  strips. The two components do not show any variation 
  along the columns or the lines of the CCD. Moreover, the 
 corrected $\left<e_1\right>$ component is well centered around zero and
does not show the systematics negative value observed by \cite{VW00}
  on CFHT data. The plots also include the cross-shape area in the
middle of the CCD where the four quadrants overlap ($X=0$ or $Y=0$).
   We cannot see significant variations 
  at those positions, which make us confident that the four 
port-readouts do not produce systematics.
}
\end{figure}

We checked the stability of this result with respect to the selection
criteria by changing the star selection or the galaxy selection
criteria. For the stars, we  changed the box-size which
encompasses the objects inside the vertical $r_h-{\rm magnitude}$ branch by 
increasing the magnitude range by 0.5 magnitude.  For galaxies, we changed
a few parameters, like the object definition (9 contiguous pixels having
0.5 $\sigma$ above the background or 6 contiguous pixels having 
1.5 $\sigma$ above the background).  We found that the amplitude 
of the variance fluctuates by $\pm 7\%$ on all scales but 2.5 arc-minute
for which the fluctuation is larger ($\pm 15\%$). These fluctuations
are within the 1-$\sigma$ error bars, therefore selection criteria do
not have a significant influence. \\
Figure \ref{shearfigmodel} shows an unusual behavior at
$\theta \approx$ 2.5 arc-minutes. The slope changes and the amplitude 
of the shear seems to increase and then to drop again on larger scales. 
This is a marginal perturbation within the error bars, but 
it turns out that it appears at the angular scale corresponding to  
one half the CCD size, so it could results from the masking 
procedure of the cross-shape area. We looked at this more
carefully by computing the shear signal inside vertical and horizontal
strips of similar width to the mask.  The results 
shown on Figure \ref{crosseffect} (which is discussed in the next Section)
do not reveal any difference between
the central regions ($X=0$ or $Y=0$), where the two segments of the 
cross are located, and the rest of the image.  We conclude
that there is no evidence for 
systematics generated by the 4-port readout 
configuration and that the increase of the signal  
is likely random error due to the lower signal-to-noise ratio on 
that scale.

\section{Analysis of systematics}
\begin{figure}
\centerline{
\psfig{figure=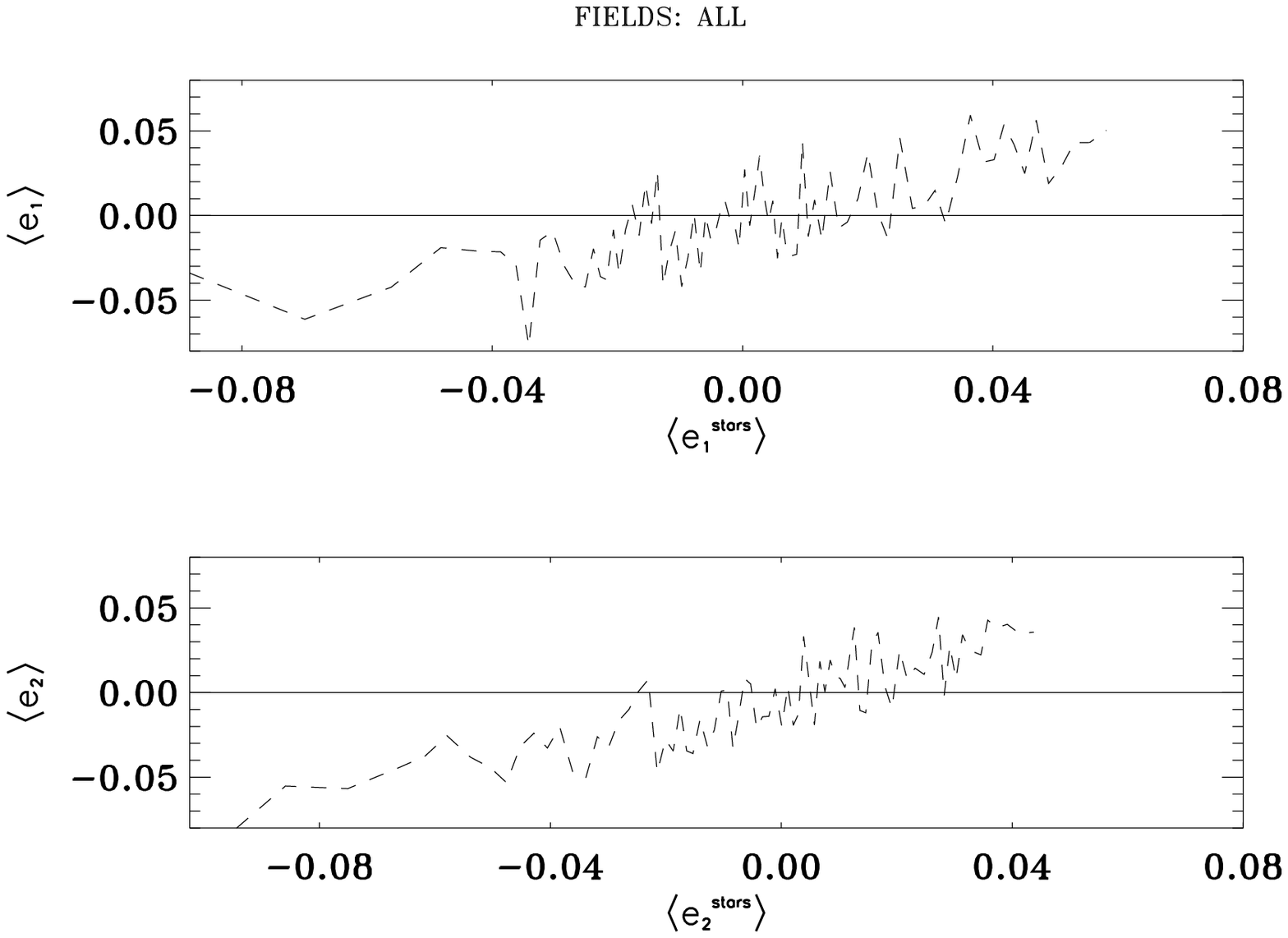,height=6cm}}
\centerline{
\psfig{figure=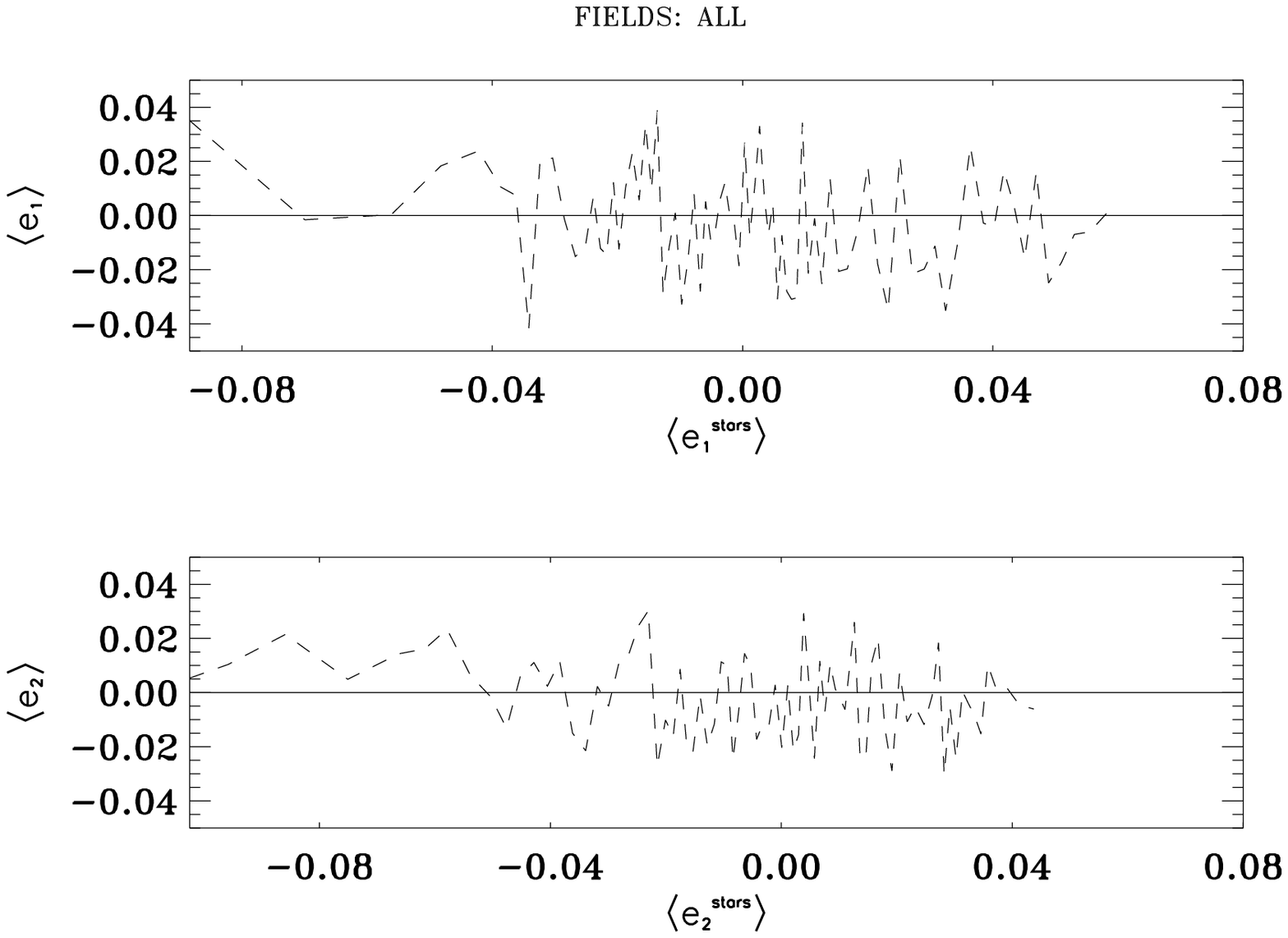,height=6cm}}
\caption{\label{correlstargal} Average galaxy ellipticity 
$\left<e_{\alpha}\right>$  compared to the average star 
ellipticity $\left<e_{\alpha}^{star}\right>$. The two plots at the top
 show the average galaxy ellipticity before correction for  the 
PSF anisotropy.  There is a tight correlation between 
  the galaxy and star ellipticity, as expected.  
  The two plots at the bottom show the same 
  components after the correction: galaxy and
star ellipticities are no longer correlated and the 
 two components of the average galaxy ellipticity are centered on 
zero.  This demonstrates that the correction works well and does not
 produce any bias.
}
\end{figure}

\begin{figure}
\centerline{
\psfig{figure=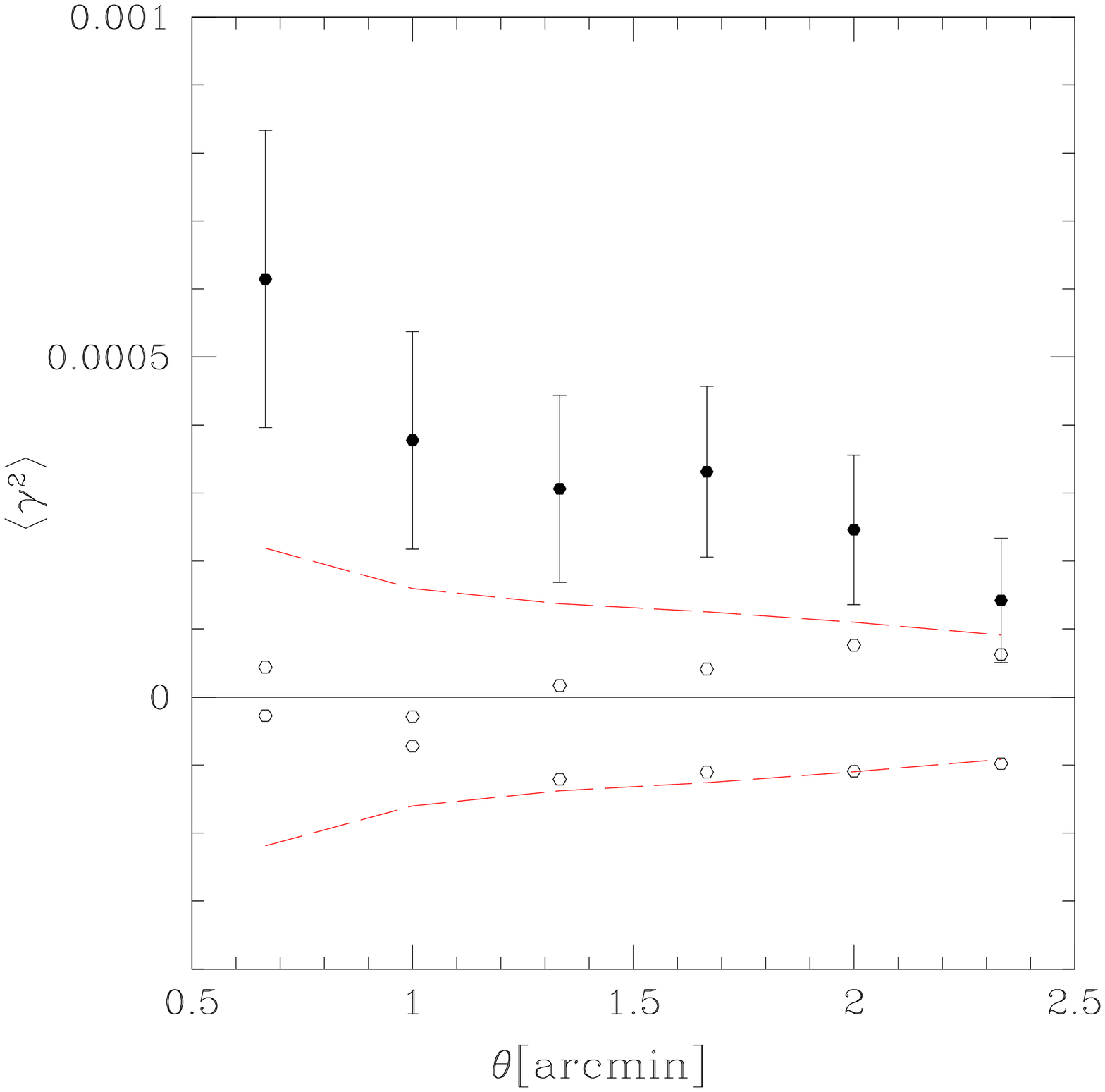,width=7cm}}
\centerline{
\psfig{figure=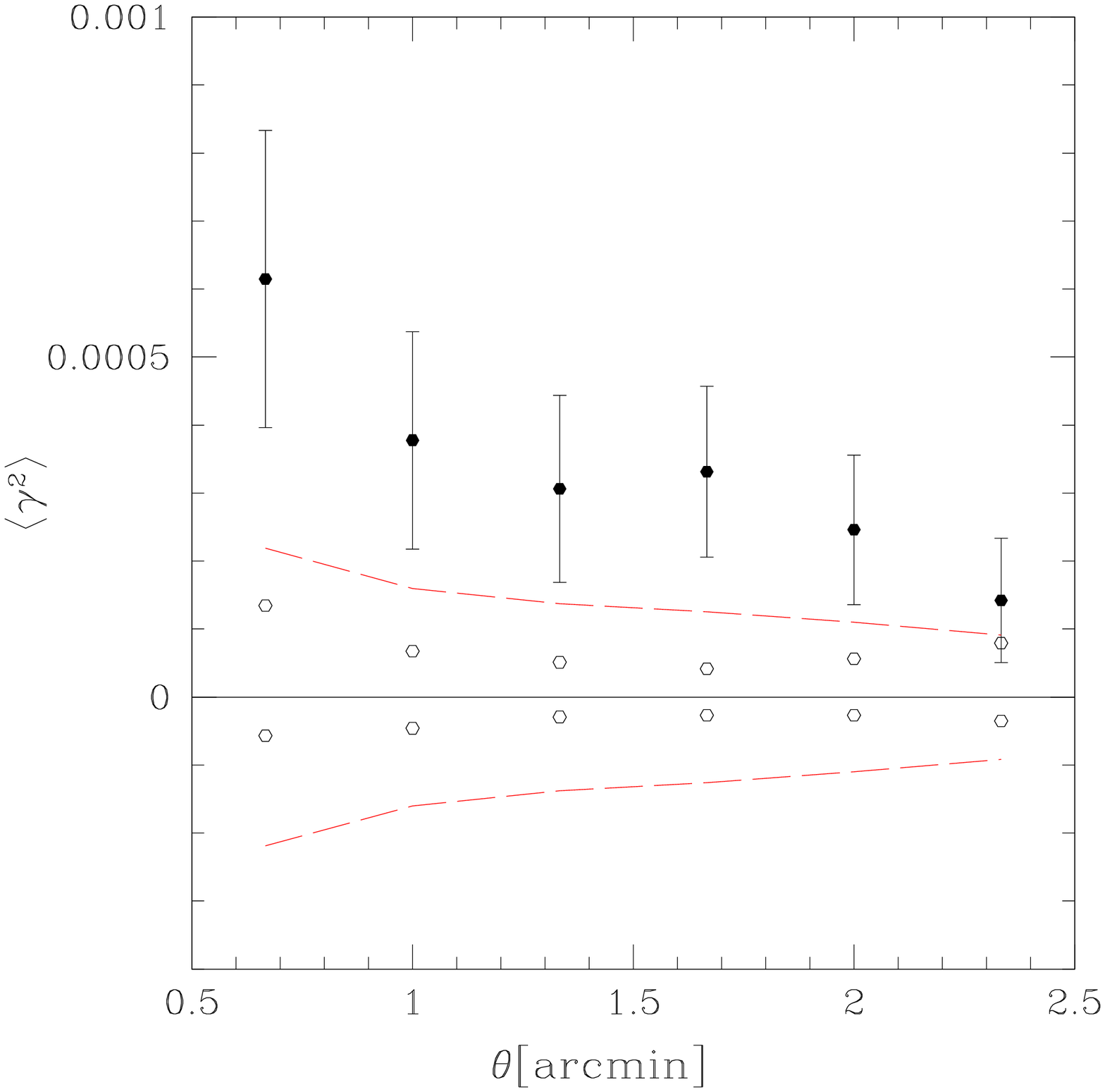,width=7cm}}
\centerline{
\psfig{figure=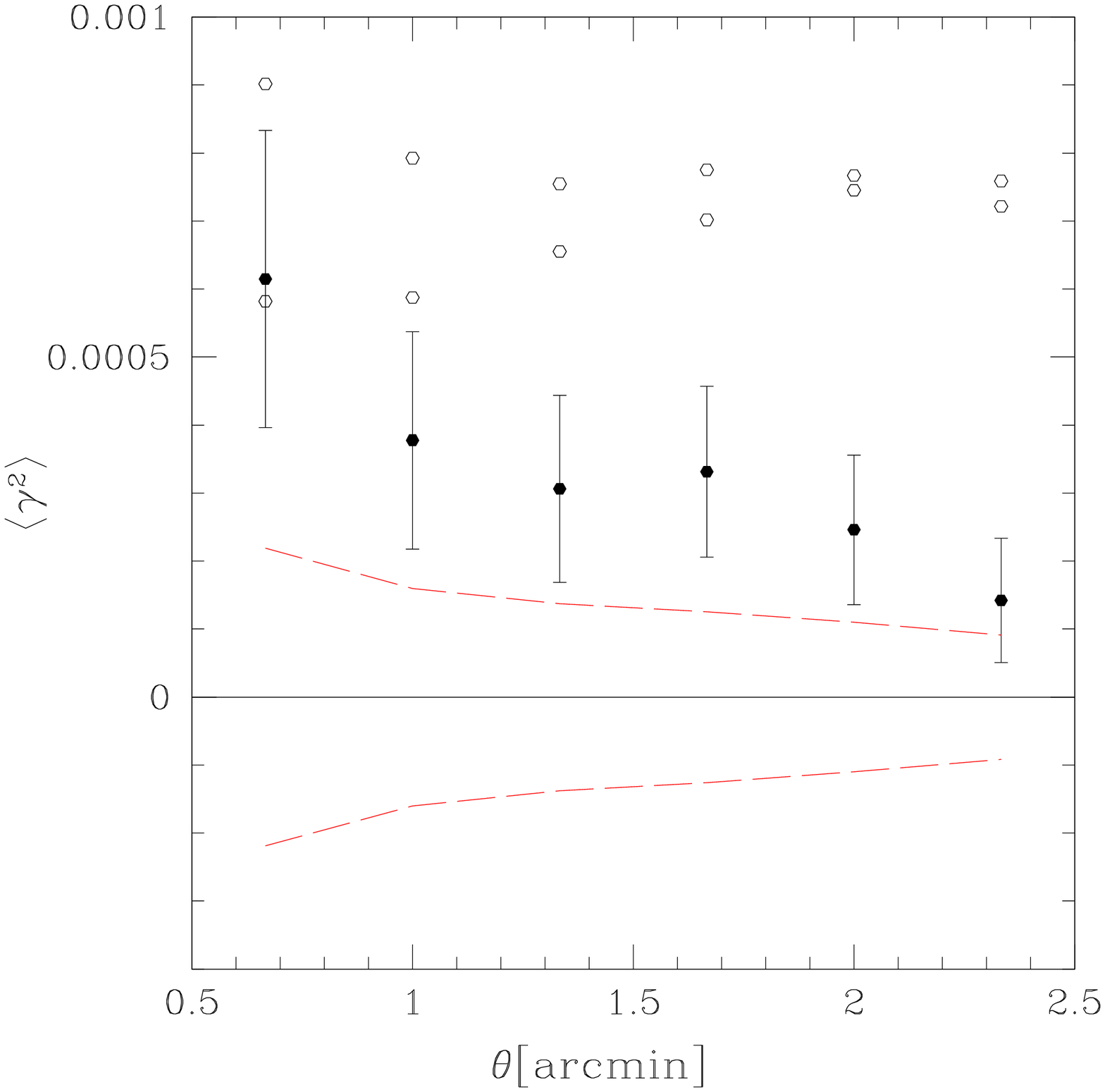,width=7cm}}
\caption{\label{summarysystematics} Analysis of possible systematics 
on the VLT fields compared to the signal. Filled circles with error
bars correspond to the cosmic shear signal; 
the dashed lines shows the 1-$\sigma$ error bars obtained from the 
100 randomizations of VLT galaxy orientations. The top panel 
shows $\left<\gamma^2\right>$ (open circles) 
measured on the galaxies sorted according to the PSF of stars
(see the two bottom panels in
Figure \ref{correlstargal}). $\left<\gamma^2\right>$ is measured
in bins of galaxies, which has a number of galaxies comparable to the number
of galaxies present in a top-hat filter of radius $\theta$.
The middle panel shows 
$\left<\gamma^2\right>$ (open circles) measured on galaxies
sorted according to their $Y$ and $Y$ location on the CCDs
(see Figure \ref{crosseffect}).  Finally, the bottom panel
shows $\left<\gamma^2\right>$ (open
 circles) as measured on galaxies
uncorrected from the PSF anisotropy (see the two top panels
in Figure \ref{correlstargal}).
}
\end{figure}

A critical issue regarding the measurement of cosmic 
shear signal is the understanding and handling of the 
 various systematics.   As in previous studies, 
  we have taken special care of this point with this new VLT data set.\\
\cite{VW00} discussed various types of systematics,  
attempted to provide some technical solutions to avoid 
part of them prior 
to use the KSB correction and provided {\it a posteriori} quality
check on the shear signal (see Sect. 5 of \cite{VW00}). 
We used similar controls for the VLT data: 

\begin{itemize}
\item {\it CCD effects:}
The spurious signal produced by bad charge transfer
efficiency, very bright stars, big galaxies or asteroids
are considerably reduced by the  masking procedure described in
Section 4. Possible residuals from charge transfer efficiency 
can be estimated by looking at the variation of the $e_1$ and 
$e_2$ components as
function of the distance of objects with respect to the readout port, for
each quadrant of the CCD. As shown in Figure \ref{crosseffect}, the
corrected components $\left<e_1\right>$ and $\left<e_2\right>$ do 
not show variations along the $X$ or $Y$ strips, except a small
negative $e_1$ component versus the $Y$ strip for $Y < 1000$. We
shall see in the following that it has a completely
negligible contribution to the variance of the shear. In fact, these
plots do not show the significant systematic residual 
of the $\left<e_1\right>$ component which was observed 
on CFHT data (\cite{VW00}). Therefore it is likely that the 
residual observed on CFHT data is intrinsic to the CFH12K 
camera. This also explains why it was not observed also by \cite{K00}
who only used UH8K data.

\item {\it residuals from the 
correction of the PSF anisotropy:}  FORS1 has a remarkable image quality 
over the field. Because of the high quality of the optical design and 
of the small field of view of FORS1 as compared to
the UH8K and the CFH12K cameras,  optical distortions are much smaller than 
in \cite{VW00}. However, active optics
on the VLT could eventually produce a new and/or unexpected 
systematic effect not fully corrected with KSB.

We tested the systematic residuals in exactly the same way as in 
\cite{VW00}.  The results are summarized in Figures
\ref{correlstargal} and \ref{summarysystematics}. The former shows that
 before the PSF anisotropy correction, the ellipticities of stars 
and galaxies are correlated. In contrast, after the correction the 
average galaxy ellipticity is zero, whatever the PSF anisotropy is.
It shows that the correction of the PSF anisotropy works very well
even in the case of active optics and does
not bias the corrected ellipticities of the galaxies.  

\end{itemize}

As pointed out in \cite{VW00}, the averaged  ellipticity of galaxies binned,
either with respect to the star ellipticity or to the CCD lines/columns
as described above, is not a strong enough test of systematics because
it is still possible that the galaxy ellipiticities 
strongly fluctuate inside very small bins. It is therefore better 
to measure the variance $\left<\gamma^2\right>$ instead of a simple
average in bins. Moreover, the result found can be compared directly
to the signal provided that, for each scale,  
the bin size is adapted  to encompass a similar number of 
galaxies as in the top-hat filter used to measure the signal.
If our corrections are  not contaminated by 
strong systematics, at all scales this variance    
must be much smaller than  the signal.\\
Figure \ref{summarysystematics} shows the results.
For each of the three panels, the filled circles with error bars show the
VLT results as shown in Figure \ref{shearfigmodel}, and the dashed lines show
the 1-$\sigma$ error bars obtained from 100 randomizations 
of the position angle of galaxies.
The open circles in the top panel show $\left<\gamma^2\right>$ measured
in bins of galaxies sorted according to the strength of the anisotropy
of stars (either $e_1^\star$ or $e_2^\star$
corresponding to the two sets of open circles). The open circles in
the middle panel show $\left<\gamma^2\right>$ with
the galaxies sorted according to the $X$ and $Y$ positions on the CCDs.
(again, each set of open circles correspond to galaxies sorted
either according $X$ or to $Y$).
It is clear from this plot that the slight negative
component $\langle e_1\rangle$ observed for $Y < 1000$ in Figure
\ref{crosseffect} is not strong enough to produce a significant systematic
compared to the signal. These two panels demonstrate that 
the residuals are always within the 1-$\sigma$ error  and fluctuate 
around zero. We conclude that these residuals are not responsible for 
the signal we detect.  \\
It is interesting to estimate how strong the systematic
effects could be if we neglected to correct the shape of the galaxies
for the PSF anisotropies.
This is shown in the  bottom panel of Figure \ref{summarysystematics}
where 
$\left<\gamma^2\right>$ is computed in the same way as for the top panel, 
but with galaxies uncorrected for the PSF anisotropy. 
The uncorrected signal shows a significant offset which is almost
insensitive to the angular scale (because the PSF is roughly constant over
each CCD). Therefore, the shape and 
the amplitude of the signal are unlikely to be produced by residuals from the
PSF anisotropy. Moreover, we see that a complete lack of anisotropy
correction produces a systematic of similar amplitude as the real signal
(and not much higher). Therefore, even a partial anisotropy correction
already permits the detection of the cosmic signal.
 
\section{Discussion}
 
The VLT data confirms the detection of a significant 
weak distortion signal on angular scales between 
0.5 to 5 arcminutes. Its amplitude and its shape 
are similar to those announced previously on these angular 
scales by four independent teams. The study of
systematics does not reveal any 
bias and reproduces similar trends as for the CFHT data but 
on a much larger sample of uncorrelated fields and on a more 
homogeneous sample (only I band, narrow seeing distribution and 
depth).

\begin{figure}
\centerline{
\psfig{figure=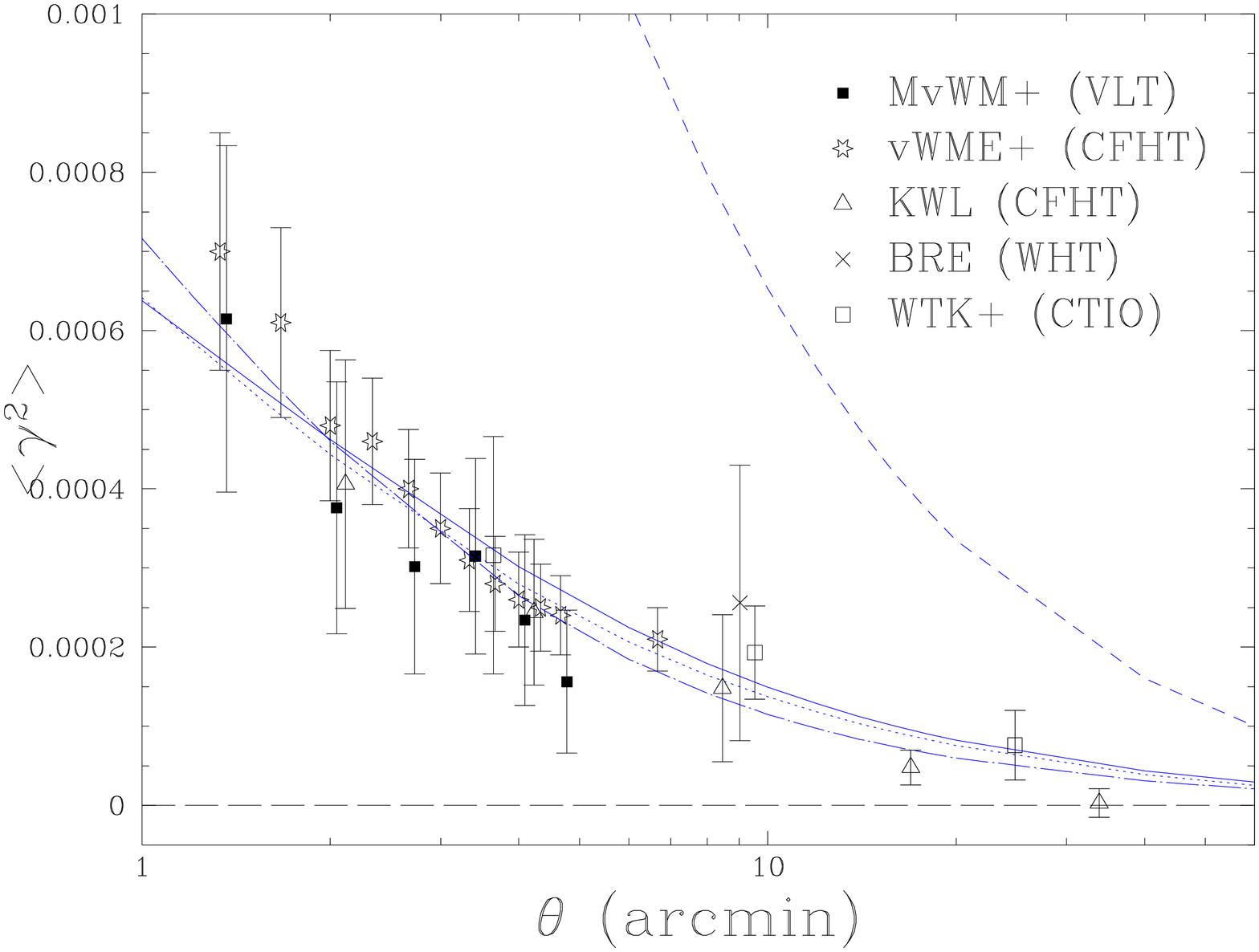,angle=270,width=9.5cm}}
\caption{\label{shearfigmodel} 
$\left<\gamma^2\right>$ as function of the
angular scale ($\theta$ is the angular diameter of a circular top-hat).
The filled squares are the VLT data presented in this work.
We also plot the previous detection: \cite{VW00} (vWME+),
\cite{BRE00} (BRE), \cite{Witt00} (WTK) and \cite{K00} (KWL).
These measurements are compared
to some current cosmological models with galaxies following the redshift
distribution Eq.(\ref{z_source}) with $z_0=0.8$. For all the models
we choose a CDM power spectrum with $\Gamma=0.21$, and
$\Omega_0=1$, $\Lambda=0$, $\sigma_8=1$  (short dash);
$\Omega_0=0.3$, $\Lambda=0$, $\sigma_8=1.02$ (dot long-dash);
$\Omega_0=1.0$, $\Lambda=0$, $\sigma_8=0.6$ (dot) and
$\Omega_0=0.3$, $\Lambda=0.7$, $\sigma_8=1.02$ (solid). The models have
been computed using the non-linear evolution of the power spectrum 
given by \cite{picdot96}.
}
\end{figure}

It is interesting to investigate what constraints on cosmological 
models we could provide from
the cosmic shear surveys completed so far by  \cite{VW00} (CFHT),
\cite{BRE00} (WHT), \cite{Witt00} (CTIO), \cite{K00}
(CFHT) and by adding the VLT data (this 
work). The five data set have been observed either in $R$ or 
in I-bands, at roughly the same depth ($I \approx 24.0$), so we can 
assume that 
the average redshifts of the sources are almost the same and should
be close to the value inferred from the deep redshift survey carried
by \cite{cohenetal}:
$z_0 \approx 0.8$ with the broad distribution given by Eq.(\ref{z_source}).
Figure \ref{shearfigmodel} shows some current 
cosmological models with the present-day data. The non-linear 
evolution of the power spectrum of density fluctuations has been 
taken into account following the prescription given 
by \cite{picdot96}\footnote{Following Peacock's advice, we do not use 
anymore the coefficients given
in his textbook, ``Cosmological Physics'' which turn out to be 
less accurate than in their paper.}.  
One can see that the cluster normalised
models fit very well the observations, at least on scales
ranging from 0.5 to 10 arc-minutes. 
However, this plot does not really illustrate the constraints
on both 
$\Omega_0$ and $\sigma_8$ we can expect from measurements of the variance of the
shear.  A more reliable study consists in exploring a  
very large set of models in a ($\Omega_0$,$\sigma_8$) space. As we can
see from Figure \ref{shearfigmodel}, and as noted
before (\cite{B97}), the dependence on the cosmological constant of the
variance of the shear is rather weak, and it is not worth to include
$\Lambda$ as a free parameter in this analysis.

Let us consider all the five cosmic shear results simultaneously.
Since they provide independent samples, we can use one single measurement
point for each of them and perform a simple $\chi^2$ minimization in
the ($\Omega_0$,$\sigma_8$) plane. From each cosmic shear measurements
we choose the point which has the best signal-to-noise,
and we avoid the large scale measures in \cite{K00} and \cite{Witt00}, as they
are likely affected by finite size effects which tend to increase the
error bars (see \cite{SC96} for a general
discussion, or \cite{B97} for a specific application to cosmic shear surveys). 
We extracted five triplets containing the
scale, the variance and the 1-$\sigma$ error,
($\theta_i$, $\gamma^2(\theta_i)$,$\delta\gamma^2(\theta_i)$), out of the
literature (the results reported on Figure \ref{shearfigmodel}) and
computed:

\begin{equation}
\chi^2={\displaystyle \sum_{i=1}^5 \left[{\gamma^2(\theta_i)-
\langle\gamma^2\rangle_{\theta_i} \over \delta\gamma^2(\theta_i)}\right]^2},
\end{equation}
where $\langle\gamma^2\rangle_{\theta_i}$ is the predicted variance for a given
cosmological model. We  computed it for 150 models inside 
the box $0 <\Omega_0<1 $ and $0.2<\sigma_8<1.4$, with $\Gamma=0.21$,
$\Lambda=0$, and $z_0=0.8$.
With $5$ data points and two free parameters, the $\chi^2$ has 3
degrees of freedom. The result is given in Figure \ref{chi2model}.
The grey scales indicates the 1, 2 and 3-$\sigma$ confidence level 
contours. The best fitted models can be described by the empirical
law:
\begin{equation}
\sigma_8\simeq 0.59 ^{+0.03}_{-0.03}\ \Omega_0^{-0.47}
\end{equation}
in the range $0.5<\theta<5$ arc-minutes. This is in remarkable agreement
with \cite{jain97} who predicted $\sigma_8\propto \Omega_0^{-0.5}$ at
non-linear scales, and this is very close to the cluster normalization
constraints given in \cite{pierpaoli} (for closed models and $\Gamma=0.23$):
\begin{equation}
\sigma_8\simeq 0.495 ^{+0.034}_{-0.037}\ \Omega_0^{-0.60} \ .
\end{equation}
This law overplotted on Figure \ref{chi2model} shows a remarkable
 agreement between these two approaches.

Our analysis is still preliminary. We have only five independent 
data points spread over a rather small angular scale. We also
assumed the peak in the redshift distribution
of the sources to be $z_0=0.8$ with the broad
distribution given in Eq.(\ref{z_source}).
This is a reasonable assumption  on the basis of the
spectroscopic survey carried out by \cite{cohenetal}, but it is still 
uncertain and needs further confirmations. We also neglected the
cosmic variance in the 
error budget of the cosmic shear sample.  Although it does not 
affect seriously the VLT data which contain 50 uncorrelated fields
and the \cite{BRE00} data point (because they estimated the cosmic
variance using a Gaussian field hypothesis),
the three other measures are probably affected. However, numerical simulations 
already indicate that cosmic variance should only 
increase our error bars by less than a factor of two, according to the
survey size (see \cite{VW00}).
We expect to have much better constraints on the redshift and 
the clustering of sources  once the VIRMOS
redshift survey will be completed (\cite{lefevretal2000}). 
On the other hand, a
measurement of the skewness of the convergence will break the degeneracy
between $\Omega_0$ and $\sigma_8$ (\cite{B97}, \cite{VW99}).

\section{Conclusion}

We have confirmed the cosmic shear signal detected on scales 
ranging from 0.5 to 5 arc-minutes using 
for the first time a large sample (50) of
uncorrelated fields obtained with the VLT UT1/ANTU.
The service mode available on this telescope gave
an unprecedented high quality and homogeneous data set. 
The fields are spread over more
than 1000 square-degrees, which minimizes the noise 
produced by cosmic variance.  The amplitude and the shape of the 
shear are similar to those measured from other 
telescopes which permits to make a strong statement about the 
robustness of the signal with respect to various sources 
of systematics.

\begin{figure}
\centerline{
\psfig{figure=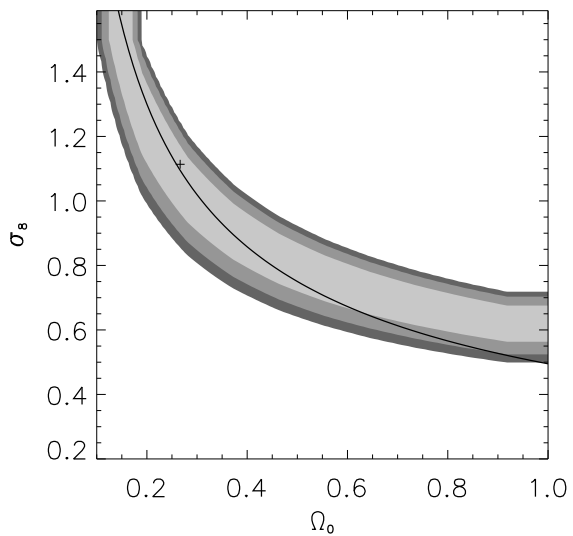,width=10cm}}
\caption{\label{chi2model} The $\Omega_0$-$\sigma_8$ constraint derived
from combined cosmic shear surveys. The three grey areas define
the 1, 2 and
3-$\sigma$ limits. The cross indicates the position of the best fit
at $\Omega_0=0.26$ and $\sigma_8=1.1$. The solid line shows
the local cluster abundance best fit (\cite{pierpaoli}).
The latter and the cosmic shear constraints
have similar shape and seem to match extremely well. 
The models have
been computed using the non-linear evolution of the power spectrum 
given by \cite{picdot96} as described in Figure \ref{shearfigmodel}.
}
\end{figure}

Assuming the signal is purely cosmic shear (that is we neglect the possible
intrinsic shape correlation), we used the four other studies published so
far to infer first constraints on cosmological models.
The cosmic shear surveys provide constraints which are in remarkable
agreement with those from cluster abundance analysis.
As compare to it, the variance of the shear is a direct
measure of the combined parameters $\Omega_0, \sigma_8$. As
soon as the residual biases are well controlled, and the redsfhit of the
sources known, the measurement of
$\langle \gamma^2\rangle$ will naturally converge to the exact
$\Omega_0, \sigma_8$ value with an accuracy that will only depends on the
accumulation of measurements on uncorrelated fields of view. This  
study shows the great potential of cosmic shear for cosmology and what 
we can expect from future wide fields surveys. In particular, 
the skewness of the convergence, which is insensitive to $\sigma_8$,
will appear as a vertical constraint
on Figure \ref{chi2model} therefore breaking 
the $\Omega_0$-$\sigma_8$ degeneracy.

{
\acknowledgements  F. Bernardeau and R. Maoli thank IAP for
hospitality were this work
has been conducted. We thank J. Peacock for clarifications about the 
use of Peacock \& Dodds' coefficients, E. Bertin, S. Colombi,
T. Hamana, D. Pogosyan and S. Prunet  for fruitful discussions
and the ESO staff in Paranal observatory for 
the observations they did for us in Service Mode.
This work was supported by the TMR Network ``Gravitational Lensing: New
Constraints on
Cosmology and the Distribution of Dark Matter'' of the EC under contract
No. ERBFMRX-CT97-0172, and a PROCOPE grant No. 9723878 by the DAAD and
the A.P.A.P.E. We thank the TERAPIX data center for providing its facilities
for the data reduction of the VLT/FORS data.  
}

\end{document}